# Succeeding at home and abroad: Accounting for the international spillovers of cities' SDG actions

Rebecka Ericsdotter Engström*[1], David Collste[2,3], Sarah E. Cornell[3], Francis X Johnson[4], Henrik Carlsen[4], Fernando Jaramillo[5], Göran Finnveden[6], Georgia Destouni[5], Mark Howells[7,8], Nina Weitz[4], Viveka Palm[6,9], Francesco Fuso-Nerini[1]

[1.] KTH Royal Institute of Technology, Department of Energy Technology, Stockholm, Sweden

[2.] Université Clermont Auvergne, Centre d'Études et de Recherches sur le Développement International, CERDI, Clermont-Ferrand, France

[3.] Stockholm University, Stockholm Resilience Centre, Stockholm, Sweden

[4.] Stockholm Environment Institute, Stockholm, Sweden

[5.] Stockholm University, Department of Physical Geography, Stockholm, Sweden

[6.] KTH Royal Institute of Technology, Department of Sustainable Development, Environmental Sciences and Engineering, Stockholm, Sweden

[7.] Loughborough University, Department of Geography, United Kingdom

[8.] Imperial College London, Centre for Environmental Policy, United Kingdom

[9.] Statistics Sweden, SCB, Department for Regions and environment, Stockholm, Sweden

* Corresponding Author. Contact e-mail: rebecka.engstrom@energy.kth.se

**ABSTRACT**

Local SDG action is imperative to reach the 2030 Agenda, but different strategies for progressing on one SDG locally may cause different 'spillovers' on the same and other SDGs beyond local and national borders. We call for research efforts to empower local authorities to 'account globally' when acting locally.

**KEYWORDS**

Spillovers, Cities, SDG interactions, localizing the 2030 Agenda

# 1. Introduction

Global achievement of the 2030 Agenda depends on local action.[1] The interconnectedness of the Agenda's Sustainable Development Goals (SDGs) and the globalized economy, however, implies that actions to advance selected SDGs in one locality create 'spillovers' that impact attainment of the same or other SDGs elsewhere.[2] Such spillovers are furthermore *strategy-dependent*[3]: different local SDG strategies, with the same local outcome, lead to different spillovers (Box 1, Figure 1).

While recent assessments of SDG interlinkages indicate that different actions for advancing one SDG may have varied effects on other SDGs[4,5], these studies do not differentiate between local and cross-border interactions.[6] Recent reports have proposed methods for national spillover accounting, as part of national SDG performance indexing.[2,7] However, these focus on mapping current and historic spillover patterns, and do not include guidance on how to consider potential future spillovers associated with new policy measures. Furthermore, they do not explore international spillovers from actions at sub-national level. 'Footprinting' and environmental life cycle assessments are abundant on project and product level[10] However, it is difficult for local decision-makers to piece such component level analyses together to build strategic sustainability plans for an entire city.

This presents a critical research gap, especially since cities' actions are often atypical of their country contexts, and many decisions that influence global development are made *below* the national level.[8] An example is climate action. The number of cities and municipalities with ambitious local decarbonisation plans is increasing worldwide.[9] While these plans are imperative to meet SDG13 (on climate action) and the Paris Agreement, assessing and managing the spillovers that they may create should also be a priority.

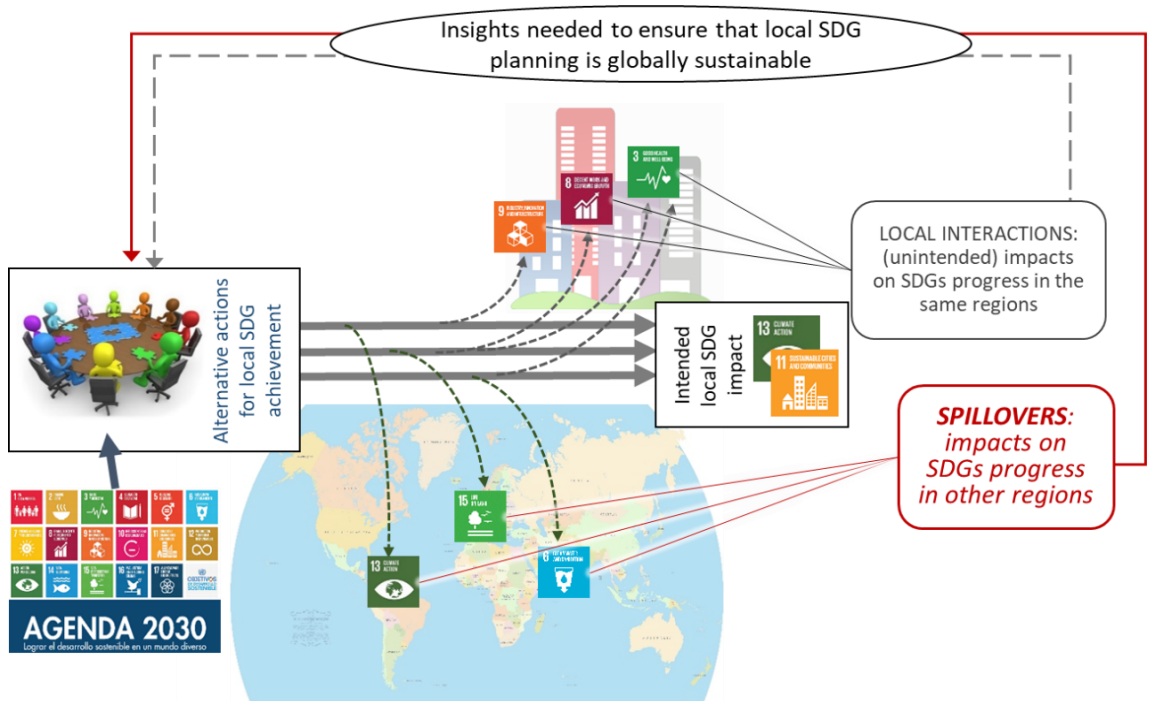

*Figure 1: Spillovers are here defined as the unintended impacts outside the geographical-political borders of the place where local SDG actions are taken. In this definition of spillovers, we do not include interactions (positive, negative, or neutral) between SDGs within the same city or municipality. These are already starting to be acknowledged in local 2030 Agenda Strategies, while spillovers to external geographies are not.*

In this comment, we define spillovers as *impacts of local SDG action beyond local and national borders that affect the ability of impact-receiving places to meet their SDGs* (Figure 1). This is a more narrow definition than proposed elsewhere.[2] We argue that addressing such spillovers when designing local SDG actions is imperative to bridge the gap between the agenda's ethos of 'global indivisibility and universality'[1] and its local (and in 17 goals and 169 targets subdivided[11]) implementation. If SDG targets can be advanced locally both *with* and *without* them, then considering spillovers (positive and negative) in local and urban sustainability strategies is an essential piece of the puzzle to reach global progress on the 2030 Agenda within this decade.

> **BOX 1: An SDG-spillover case study from southern Sweden**
>
> The municipality of Oskarshamn, Sweden, aims to become climate neutral by year 2030. This requires decarbonizing local transport, industry, agriculture and energy use in buildings.
>
> A study[3] of the potential impacts on water and land use from various decarbonisation pathways showed that increased electrification of the local society would increase water and land use within Sweden, while increased local use of biofuels would increase water and land use abroad. However, all strategies met the local policy makers goal of climate neutrality.
>
> Translating this to impacts on SDGs, one strategy to decarbonize Oskarshamn and advance SDG 7 (on clean energy) and SDG 13 (on climate action) could impede progress on SDG15.1 (on terrestrial biodiversity) in Sweden, while another strategy might compromise SDG 6.4 (on sustainable water use) in Germany or Brazil.
>
> The study reveals that spillovers are largely strategy-dependent, both in magnitude and in their geographical distribution. Furthermore, the geographical distribution of spillovers matter. For the example of water related spillovers, humid and arid regions differ greatly in their vulnerability to enhanced consumptive use of water.

## 2. The challenge: Moving spillovers from invisible and uncertain, to reflected and assessable

Several challenges are present for spillover accounting at the local level. First, mandates for local Agenda 2030 decision-making focuses on local SDG attainment. Formally, local decision-makers stand unaccountable for the spillovers of their decisions, and incentives to look beyond the interests of local voters are hard to find. Second, delivering on existing local mandates is challenging enough in most places, with commonly limited financial and human capacity and resources of local authorities.[12] Third, even where there is will and resources, it is inherently difficult to assess the potential future spillovers from local SDG action. Uncertainties are unavoidable, particularly when assessing future impacts.[3]

We take a Swedish perspective here to highlight the challenge of spillovers from sub-national level. The country is considered one of the front-runners in SDG attainment[2], and holds a long-standing high profile in international cooperation. In 2003, Sweden adopted a national Policy for Global Development (PGU – "Politik för global Utveckling") with the core principle that all political decisions made in Sweden should contribute to a sustainable global development.[13] PGU has since been incorporated in Sweden's formal adoption of the 2030 Agenda.[13] Meanwhile, many Swedish municipalities and cities aim to contribute to the national achievement of the 2030 Agenda. However, there are no clear procedures for Swedish local governments to live up to the principles of PGU in their decision-making.[14]

While the Swedish case is not universal, many cities around the globe have high ambitions to progress on the 2030 Agenda, but similarly lack tools to assess and manage potential spillovers. And with less than a decade left to transform our societies, differentiated action is increasingly necessary.

**3. A research agenda to enable spillover consideration in local SDG strategies**

Figure 2 was used as starting point for the authors' deliberative exploration of spillover challenges, which lead to the following research recommendations.

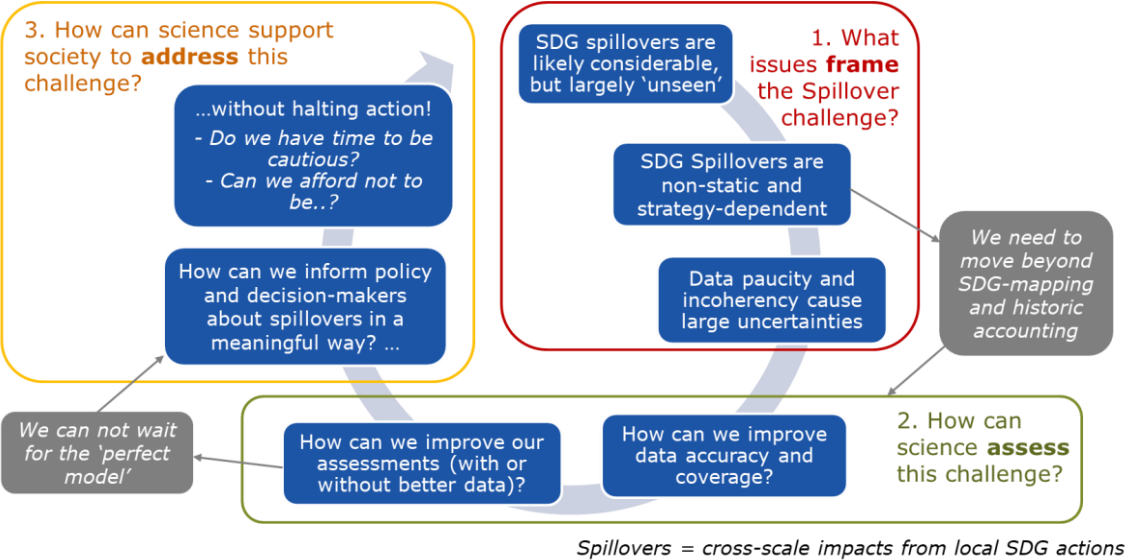

*Figure 2: Key questions to structure discussion of the spillover challenge for science and society.*

*3.1. Fit spillover assessment approaches to local realities*

Decision-makers need analyses that are specific to their decision contexts in order to make spillover assessments feasible and actionable. Approaches for spillover considerations must therefore be built to

- fit **existing capacity** of local governments. This includes developing user-friendly, transparent tools and interfaces that make spillover consideration *practical* across diverse expert groups and with lay users.[15] For this, good information visualizations may be as important as improved quantitative evaluations. Additionally, any process for 'rapid assessment' of spillovers must provide decision-support that embraces uncertainty and avoids the dangers of over-simplification.

- be guided by and flexible to how SDG spillovers play out in **the local case** – with current social, economic and environmental development levels**.** Participatory approaches for untangling spillovers (and testing how they can be handled) are important here. Such approaches are tried-and-tested ways to facilitate learning and deal with diverse local contexts but are still poorly integrated into the larger-scale assessments that dominate the SDG research field.[6]

- recognise the role of established **local processes**, planning paradigms and organizations. Building spillover assessments in line with local planning practices and identifying when, where and how spillovers should be considered are essential for effective translation of research into urgently needed practice.

Development of such practical processes should take inspiration from established methods for science-policy-society interaction in integrated assessments (Table 1).

Given the strategy-dependence of spillovers, assessments must enable evaluation of alternative SDG pathways, comparing impacts on directly targeted vs. indirectly affected SDGs and facilitating multi-perspective evaluation of local vs. global and near-term vs. long-term priorities.

*Table 1. Examples of established methods for science-policy-society interaction in integrated assessments.*

| **Participatory methodology or process** | **Example References** |
| --- | --- |
| *Integrated Water Resource Management* | Global Water Partnership (GWP). *Integrated Water Resources Management*. Global Water Partnership (GWP) Technical Advisory Committee (2000). |
| *Transboundary nexus methodology* | de Strasser, L. *et al.* A Methodology to Assess the Water Energy Food Ecosystems Nexus in Transboundary River Basins. *Water* **8**, 59 (2016). |
| *Three Horizons for SDGs* | Collste, D. *et al. Three Horizons for the Sustainable Development Goals: A Cross-scale Participatory Approach for Sustainability Transformations.* (2019) doi:10.31235/osf.io/uhskb. |
| *Visions of a good Anthropocene* | Pereira, L., Hichert, T., Hamann, M., Preiser, R. & Biggs, R. Using futures methods to create transformative spaces: visions of a good Anthropocene in southern Africa. *Ecology and Society* **23**, (2018). |

*3.2. Advance, coordinate and make accessible: tools, data and uncertainty*

The envisioned local spillover assessments rely on the availability of interchangeable tools, subject to topic of analysis, data availability, time and resources. Since 2015, the scientific community has presented a plethora of tools and methods for assessing how SDGs can be met and how they relate to each other (Table 2). Many build on sustainability assessment tools that have evolved over decades. With their evolution, model complexity, data demands, and computational time have grown. Different tools study different objects with different system boundaries and, as such, they are not easily comparable or combinable.[16] To enable spillover analysis of local actions, beyond pilot cases and demo-cities – and to facilitate meaningful comparison between approaches – we call for the following strategic research coordination priorities.

- Increase international collaboration on data gathering, sharing and transparency to improve accessibility of information on SDG spillovers. This should build on the FAIR data principles.[17]

- Encourage multi-model coordinated assessments, taking inspiration from, for example, the Coupled Model Intercomparison Project (CMIP) aimed at improved understanding of climate changes in a multi-model context.[18] This should include aligning scopes and defining common baseline scenarios and key assumptions across models, while allowing for variations in resolutions and dynamics. Comparing results can then enrich the analyses and offer complementary perspectives.

- Increase capacity to highlight and handle uncertainty in modelling. Beyond being unavoidable, uncertainties can inform us about qualities of the systems we analyse and give important insights to improve adaptive management.[19]

As with all ambitious future-casting modelling efforts, it will be important to learn from past attempts (and failures) to project the future with models. Acknowledging elements that are quantifiable only to some degree and have considerable remaining uncertainties, such as non-linear dynamics of system resilience, is also important to help avoid over-simplification or partial optimizations. To be effective, these efforts should be shared in open scientific knowledge hubs (building on or inspired by e.g. [18,20]) where data on SDG interactions, tools and problem tackling processes can be uploaded, discussed and reconciled.

### *3.3. Guide prioritisation*

Consideration of spillovers must be made possible without overwhelming local planners and decision-makers in never-ending cross-border, cross-SDG impact analyses. This demands that the international research community reconciles current knowledge and guides analysis to where it is most impactful. An important, third effort is therefore to review and identify SDG *targets*, *actions* and *localities* that are

- most likely to create the largest spillovers and
- most vulnerable to receive spillovers.

Clues to identifying these are found in recent research, reports and the 2030 Agenda itself. The vulnerability of least developed countries (emphasised throughout the Agenda) is one example. The importance of achieving SDG12 (on sustainable production and consumption) in wealthier countries to enable achievement of this and other SDGs globally is another.[21] Assessments of the environmental footprint of Swedish consumption has shed light on the latter.[22]

The current global pandemic has highlighted the vulnerability of developed and developing countries alike to globalised supply chains, which may encourage more localised production and consumption. Although this might support SDG12, it could also lead to negative spillovers through higher prices and reduced access to crucial products for least developed countries with minimal industrial capacity.

A consolidation of recent mappings of SDG interactions and spillover evaluations (including, for example, work of [7,23,24]) should also help highlight the goals and targets that largely condition the whole Agenda. We anticipate that progress on, for example, poverty eradication (SDG1), sustainable energy access (SDG7) and non-violence (included in SDG16) are prerequisites for local progress on most other SDGs. Verifying this, and identifying additional 'overriding targets', could help pinpoint key spillovers that either halt or advance their progress (and hence should be minimized or maximized).

Lastly, we call for prioritisation of spillovers in established accounting procedures. We have to date found no indicators for local SDG monitoring that captures spillovers. For example, indicators used in cities and municipalities around Sweden are chosen based on local data availability and, as a consequence, do not capture non-local impacts.[14] Formal SDG indicators[11] with strong correlation to spillovers, together with recently proposed spillover indicators[2], should be highlighted and prioritized in local, national and global progress reporting.

## 4. A call for expanded responsibilities

Our comment is a call for research that enables and empowers local decision-makers to account for spillovers from local SDG implementation strategies. Many cities with large and varied spillovers have high ambitions to contribute to global sustainability. Availability of tools, guidance and incentives are vital to enable them to address their spillovers – and fully meet those ambitions. We believe the above-mentioned research calls can deliver critically missing tools and guidance.

To enhance incentives, we encourage the formation of an international community of local decision-makers and cities that account for their spillovers. Bringing experiences and ambitions together would raise the profile of spillover management and promote ethical dimensions of the 2030 Agenda at the local level. It could also help expand the mandates of local decision-makers. The pledge to "leave no one behind" means that this community need not be restricted to elected local government officials but may also incorporate other local actors.

However, addressing spillovers at the local level is not a silver bullet for effective SDG implementation. International diplomacy, cooperation and development finance are also crucial to mitigate negative spillovers.[2] The efforts to address local spillovers that we call for here must therefore be accompanied by ongoing work to:

- Strengthen the authority and ambitions of international negotiations and treaties on sustainable and fair development. This includes strengthening the implementation process of the 2030 Agenda as a whole.

- Support new international partnerships that improve institutional capacity in least developed countries and incorporate norms of good governance, in line with SDG16 and SDG17.[25] This includes delivering on development support targets already embedded in the agenda.[7]

Spillover consideration, in the form discussed here, may seem like a cumbersome add-on that only risks delaying local action on the 2030 Agenda. However, we believe that every effort to manage spillovers will also improve management of local SDG interactions. Furthermore, a focus on spillovers in local SDG implementation could strengthen the Agenda's ethos of universality and indivisibility 'on the ground'. Time is short, and all local SDG actions should be encouraged. Still, we must ensure that local advancements do not cripple global progress.

*Table 2: Common tools for assessing SDG interaction. Not comprehensive list.*

| Type of assessment | Name of example model | SDG Focus | Boundaries | Examples of published studies touching on spillovers | Strengths of tool to capture Spillovers | Limitations of tool to capture Spillovers |
|---|---|---|---|---|---|---|
| LCA (Life Cycle Assessment) | ISO 14040, and ISO 14044 (2006 a and b) | SDGs 2, 3, 6, 7, 11, 12, 13, 14, 15 | The whole value chain should be included without spatial or temporal boundaries. | (Knobloch et al. 2020; Borggren et al. 2013; Sen et al. 2020) | Useful to capture problem-shifting including geographical spillovers since the whole lifecycle, including impacts in other places and times, are assessed. | Does normally not include site-specific data, which may make it difficult to analyze where impacts occur. Lack of relevant data may limit conclusions. |
| Social LCA (S-LCA) | Guidelines for social LCA (Benoît et al. 2010) | SDGs 1, 2, 3, 4, 5, 8, 10, 11, 12, 16, 17 | The whole value chain should be included without spatial or temporal boundaries. | (Ekener-Petersen, Höglund, and Finnveden 2014) | S-LCA is a useful to tool to capture spillovers since the whole value chain regardless of geographical location should be included. | Lack of relevant data may limit studies and conclusions. |
| Multi-Regional Input-Output Analysis (MRIOA) | Exiobase (Stadler et al. 2018) | SDGs 1, 2, 3, 6, 7, 8, 9, 10, 11, 12, 13, 14, 15, 17 | Can cover all nations or regions worldwide. Data are typically for a specific year. | (Sen et al. 2020; Wood et al. 2020) | MRIOA analyses in which countries impacts occur and can therefore identify spillovers. | Data are aggregated for product groups and countries/regions. |
| System Dynamics Models | iSDG, International Futures | Targets from all SDGs incorporated | For national and regional decision-making. Time horizon: 20-30 years. Includes social, economic and environmental sectors. | (Pedercini et al. 2019) | Integrates all SDGs, combines qualitative maps with quantitative modelling inviting stakeholder consultation, and opens the scope for integrating spillovers. | Typically not incorporating cross-border interactions in current applications, but can be integrated. |
| Water-Energy-Food Nexus modelling | CLEWs (Climate Land Energy Water systems) assessments | SDGs 2, 6, 7, 13 and 15 to varying extent | Global, trans-boundary, national and local scopes. Time-horizon: 10-50 years. Land/food, water, energy and climate policies. | Trans-boundary: (Almulla et al. 2018). Local: (Engström et al. 2017; 2019). Global: (United Nations 2014). | SDG interactions in CLEW sectors are captured. Spillovers across transboundary waters captured in detail. Recent review of water and land-use footprints of energy use highlight local spillovers | Spillovers not directly linked to the CLEW sectors are not assessed. The approach is quantitative, bottom-up and future-focused. This makes data scarcity a limiting factor in analysing spillovers. |
| NDC-SDG interaction assessment | Network analysis and prioritization frameworks | The whole Agenda | Boundaries are concept-specific depending on type of interactions addressed. | (Jaramillo et al. 2019; Weitz et al. 2018; Allen et al. 2019; Nerini et al. 2019) | Easily adaptable to contexts and applications. Qualitative information that is standardized | Spatial dimension, difficult to quantify, without coupling to other types of models |

**Data availability Statement**

No datasets were generated or analysed during the current study

**Acknowledgements**

In addition to the authors, valuable contributions to this paper was made by Björn Hugosson, Head of the Climate Unit at the City of Stockholm's Mayor's Office, who participated in the explorative workshops that formed the core ideas of this piece.

The authors further thank Erica-Dawn Egan, Karin Larsdotter and KTH Sustainability Office for support in arranging the workshops that lead to the here expressed insights. We also thank Jennifer Castor and Kaylyn Bacha for assisting in the workshop and its documentation.

R.E. Engström and F. Fuso Nerini acknowledge funding from the Swedish Research Council Formas, grant number 2018-01253. S.E. Cornell acknowledges partial support from the European Research Council under the EU Horizon 2020 research and innovation programme (grant agreement 743080 – ERA). H. Carlsen and N. Weitz acknowledge partial support from MISTRA—The Swedish Foundation for Strategic Environmental Research through the research programme MISTRA Geopolitics.

**Author Contributions**

R.E.E. and M.H. developed the initial conceptualization and hypothesis for this paper. All authors contributed with substantial input to the core content of this paper through explorative workshops, each bringing a unique perspective and experience of the issue discussed.  R.E.E., D.C. and F.F.N. wrote the first draft of the text. All authors reviewed and actively contributed to the writing, review and editing of the final version of the text. D.C., R.E.E, G.F., F.J. and F.F.N. provided the content of Table 2. R.E.E., G.F., M.H. and G.D. contributed with project administration.

**Competing interests**

The authors declare that there are no competing interests.